\documentclass[review]{elsarticle}

\usepackage{amssymb,bbm}
\usepackage[tbtags]{amsmath}
\allowdisplaybreaks
\usepackage{multirow}

\usepackage{lineno,hyperref}
\usepackage{threeparttable}
\usepackage{placeins}
\modulolinenumbers[5]

\journal{Pattern Recognition Letters}

\begin{document}

\begin{frontmatter}

\title{Factor Decomposed Generative Adversarial Networks for Text-to-Image Synthesis}


\author[1]{Jiguo Li}\ead{jgli@fudan.edu.cn}
\author[2]{Xiaobin liu}\ead{xbliu@tecent.com}
\author[1]{Lirong zheng\footnote{Corresponding Author: lrzheng@fudan.edu.cn}}\ead{lrzheng@fudan.edu.cn}

\affiliation[1]{organization={School of Information Science and Technology, Fudan University},
            city={Shanghai},
            postcode={200433}, 
            country={China}}
\affiliation[2]{organization={Tecent.Inc},
            city={Beijing},
            country={China}}


\begin{abstract}
Prior works about text-to-image synthesis typically concatenated the sentence embedding with the noise vector, while the sentence embedding and the noise vector are two different factors, which control the different aspects of the generation. Simply concatenating them will entangle the latent factors and encumber the generative model.
  In this paper, we attempt to decompose these two factors and propose Factor Decomposed Generative Adversarial Networks~(FDGAN). To achieve this, we firstly generate images from the noise vector and then apply the sentence embedding in the normalization layer for both generator and discriminators. We also design an additive norm layer to align and fuse the text-image features. The experimental results show that decomposing the noise and the sentence embedding can disentangle latent factors in text-to-image synthesis, and make the generative model more efficient. Compared with the baseline, FDGAN can achieve better performance, while fewer parameters are used.
\end{abstract}


\end{frontmatter}


\section{Introduction}
%
%
%
%

It is a challenging problem to synthesize realistic images from the visual text descriptions, which has many potential applications, such as computer-aided design, human-computer interaction, image editing, \textit{etc}. It has been an active topic since Reed \textit{et al.}~\cite{reed2016generative} firstly synthesized images with a resolution of 64$\times$64 from the text description using generative adversarial networks~(GANs)~\cite{goodfellow2014generative,mirza2014conditional}. Remarkable progress has been achieved in recent years~\cite{zhang2017stackgan,zhang2018stackgan++,xu2018attngan} to synthesize more realistic images with higher resolutions.

\begin{figure}[t]
  \centering
    \includegraphics[width=.99\linewidth]{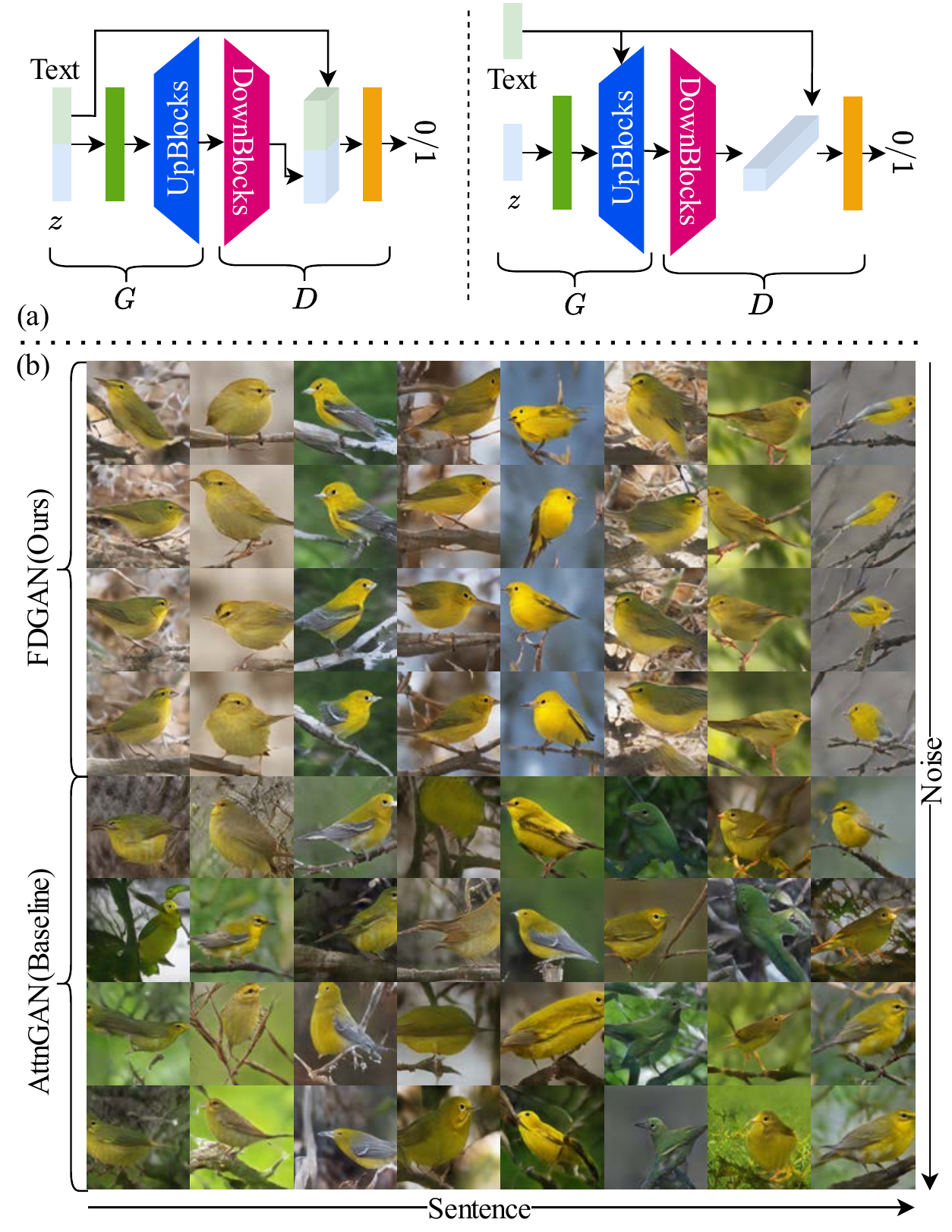}
\caption{(a) The comparison of prior typical framework~(left) and our proposed FDGAN~(right). The typical framework concatenates the sentence embedding with the noise vector. Our proposed FDGAN decomposes the sentence embedding and the noise vector. (b) Factor decomposition generation results of our proposed FDGAN~(top 4 rows) and AttnGAN~(bottom 4 rows) on CUB 200 dataset. AttnGAN, which concatenates the noise and the sentence embedding, entangles the latent factors. While our proposed FDGAN can disentangle the latent factors by decomposing the noise vector and the sentence embedding. (Best viewed in color.)}
\label{fig:factor_decomposed}
\end{figure}

Priors works~\cite{zhang2017stackgan,xu2018attngan,zhu2019dm} typically concatenated the sentence embedding, which is extracted from the text description, with the noise vector, which is sampled from the normal distribution. However, the sentence embedding and the noise vector correspond to different aspects of the generated image. When the image mainly contains only one object, the text usually describes the texture of the object, while the noise should correspond to the pose, \textit{etc.} When the image contains multiple objects, the text typically describe the scene and the relationship of the multiple objects, while the noise should correspond to the details of the object. Hence, the sentence embedding and the noise are two different factors. Simply concatenating them will entangle the latent factors, which will hurt the text-to-image synthesis, as shown in Fig.~\ref{fig:factor_decomposed}. A better method that can decompose the sentence embedding and the noise is expected for optimizing the text-to-image synthesis framework.

Factor decomposition has been proposed in other related tasks to design a more interpretable and efficient model. In video generation, MocoGAN~\cite{tulyakov2018mocogan} decomposed the object motion and the content by sampling different random noise for motion and content. In image style transfer, the image is typically decomposed into the content and the style~\cite{huang2017arbitrary,liu2019few}. So the image content can be kept when the style is transferred. We argue that in text-to-image synthesis, the sentence embedding and the noise vector should also be decomposed because they correspond to the different aspects of the generation. 

\FloatBarrier 

Based on the above discussion, in this paper, we aim to decompose the noise and the sentence embedding in text-to-image synthesis. To achieve this, we (1) separate the sentence embedding from the sampled noise; (2) align the text embedding feature with the image intermedia feature and apply the aligned feature as the bias for the normalization layer; (3) apply the sentence embedding as the bias of the normalization layer for the conditional discriminator. We also design an additive instance norm layer to align the text-image features and apply the sentence embedding into the generator and discriminators. The difference between our FDGAN and prior works, as well as the factor decomposition generation results, are shown in Fig.~\ref{fig:factor_decomposed}. Compared to AttnGAN, which is a widely used baseline, our FDGAN can disentangle the latent factors better, which correspond to the noise and the sentence embedding, respectively.
Summarily, our contributions are three-fold:
\begin{enumerate}
    \item We propose a novel framework, FDGAN, to decompose the two main factors in text-to-image synthesis: the sentence embedding and the sampled noise. Experimental results show that FDGAN succeed to disentangle the latent factors in text-to-image synthesis. 
    \item  We design the additive instance norm~(AddIN), which is a variant of AdaIN~\cite{huang2017arbitrary}, to apply the sentence embedding into both the generator and the discriminators.
    \item Extensive experiments are conducted to demonstrate the effectiveness and the efficiency of the proposed FDGAN. The experimental results show that FDGAN can achieve better performance with fewer parameters. To the best of our knowledge, FDGAN is the first model that both the performance and the computational efficiency are improved than AttnGAN~\cite{xu2018attngan}. 
\end{enumerate}

\section{Related Works}
\subsection{Factor Decomposition in Style Transfer}

Style transfer aims to manipulate the style of an image while preserving the content under the assumption the image can be decomposed into two parts: content and style. 
Adaptive instance normalization~(Adain)~\cite{huang2017arbitrary} is a widely used technology to synthesize stylized, realistic, and diverse images~\cite{lee2018diverse,liu2019few}.
Style transfer can be modeled as an intra-modality conditional generation task, while text-to-image synthesis is a cross-modality conditional generation task. Motivated by the similarity of these two problems, we attempt to decompose the factors in text-to-image synthesis and apply the conditional information via instance normalization.

\subsection{Text-to-Image Synthesis}
Text-to-image synthesis aims to generate realistic images from the given text description. Reed~\textit{et al.}~\cite{reed2016generative} firstly applied deep based methods to text-to-image synthesis, generating images with a resolution of $64\times64$ by generative adversarial networks~(GANs)~\cite{goodfellow2014generative}. The following works~\cite{zhang2017stackgan,zhang2018stackgan++,qiao2019mirrorgan} improved the resolution of the synthesized images to $256\times256$ by stacking multiple generation stages and generating images with increasing resolutions. Xu~\textit{et al.}~\cite{xu2018attngan} proposed AttnGAN to exploit the word level fine-grained description in the text by learning the cross-modality attention between the word features and the local image features, synthesizing fine-grained images from the input text descriptions. 

Based on AttnGAN, SEGAN~\cite{tan2019semantics} and SDGAN~\cite{yin2019semantics} proposed to optimize the feature representation via contrastive loss; DM-GAN~\cite{zhu2019dm} refined the synthesized image further with an additional dynamic memory module; ReFiGAN~\cite{cheng2020rifegan} enriched the text embedding by matching the semantically similar captions and generated images from these enriched descriptions. However, all these improved methods (1) entangled text condition with the noise in the generation; (2) brong additional parameters~(\textit{e.g.} dynamic memory for DMGAN, text retrieval for RiFeGAN) or computation~(\textit{e.g.} contrastive loss for SEGAN and SDGAN). By comparison, our FDGAN can achieve better performance with almost no additional computation and much fewer parameters by factor decomposition.

\begin{figure*}[t]
    \centering
      \includegraphics[width=0.99\linewidth]{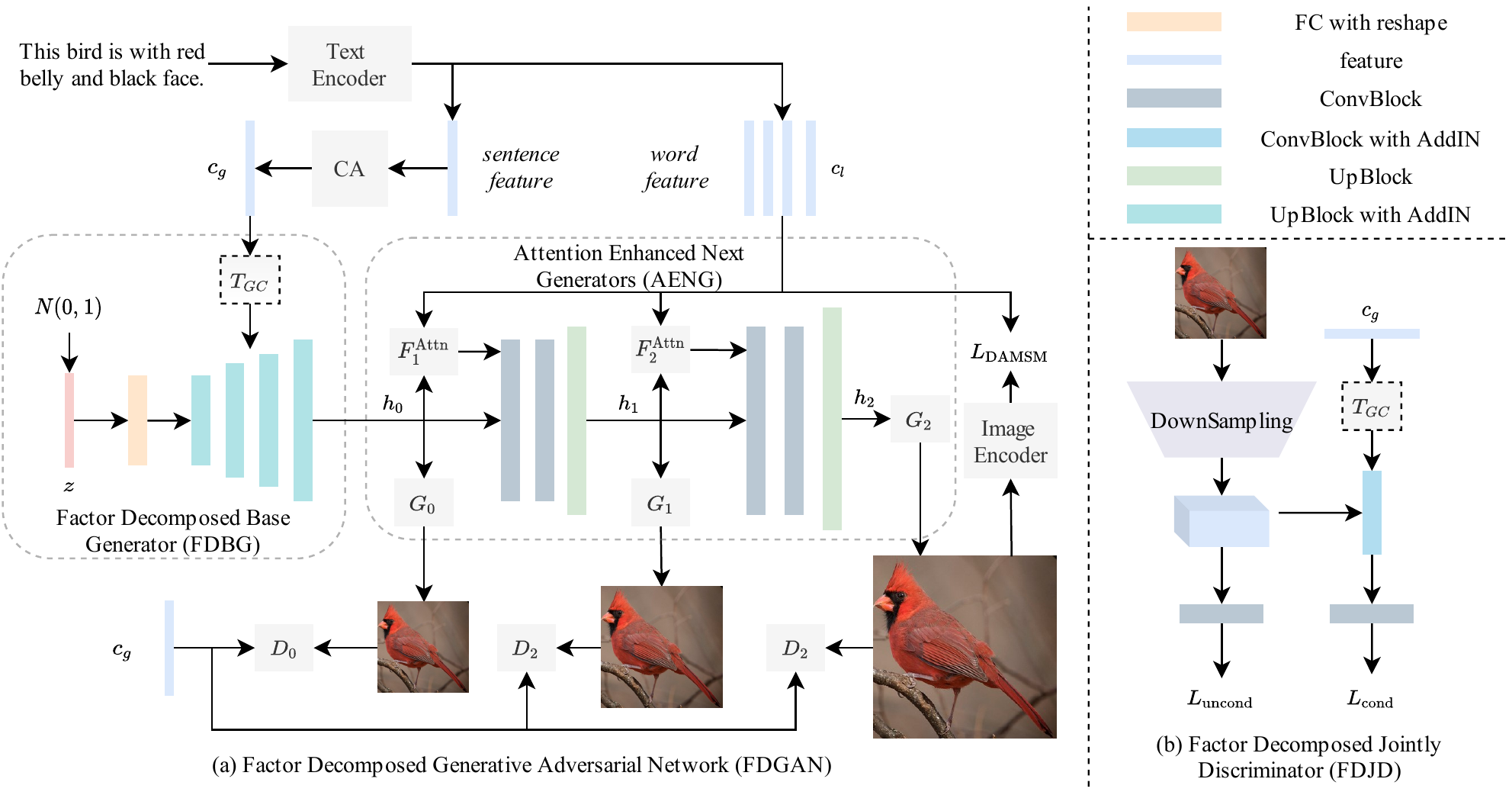}
  \caption{Framework of Factor Decomposed Generative Adversarial Network~(FDGAN). (Best viewed in color.)}
  \label{fig:framework}
  \vspace{-2mm}
\end{figure*}

\section{Factor Decomposed GAN}

\subsection{Framework}
Based on the idea that the sentence embedding and the noise should be decomposed, we propose Factor Decomposed Generative Adversarial Network~(FDGAN). Following~\cite{zhu2019dm}, we use a stacked generator with attention modules~\cite{xu2018attngan} as the backbone. The RNN based text encoder and Inception-V3 based Image Encoder are pretrained to learn the cross-modality attention. As illustrated in Fig.~\ref{fig:framework}, FDGAN consists of a multi-stage generator and several discriminators for modeling image distribution at increasing resolutions. The generator has two submodules: Factor Decomposed Base Generator~(FDBG) and Attention Enhanced Next Generators~(AENG). In FDBG, we decompose the global condition $c_g$ and the noise vector $z$, applying the global condition in the upsampling block after a transformation block $T_{GC}$. Similarly, the global condition $c_g$ is also applied in the discriminator to model the conditional distribution. To bring the gap between the text feature and the image intermedia feature, we design the \textbf{Add}itive \textbf{I}nstance \textbf{N}ormalization layer~(AddIN), which is a variant of AdaIN~\cite{huang2017arbitrary}.
\subsection{Additive Instance Norm}\label{sec:addin}
To modulate the global condition into the image generation in an global manner, inspired by the prior works about condition normalization~\cite{dumoulin2016learned,huang2017arbitrary,yin2019semantics} we propose the \textbf{Add}itive \textbf{I}nstance \textbf{N}orm~(AddIN).

\noindent\textbf{Instance Norm:} Given an input batch $x\in \mathbb{R}^{B\times C \times H \times W}$ and a condition feature $y\in \mathbb{R}^{B\times C_y}$, IN normalizes the mean and the deviation for each individual feature channel:
\begin{align}
\text{IN}(x) = \gamma\left(\frac{x-\mu(x)}{\sigma(x)}\right) + \beta,
\end{align}
where $\gamma,\beta \in \mathbb{R}^{B\times C}$ are learnable affine parameters which can be derived from data, $\mu(x)$ and $\sigma(x)$ are calculated across spatial dimensions independently for each channel and sample:
\begin{align}
    \mu_{nc}(x) = \frac{1}{HW}\sum_{h=1}^H\sum_{w=1}^W x_{nchw},
\end{align}
\begin{align} 
    \sigma_{nc}(x) = \sqrt{\frac{1}{HW}\sum_{h=1}^H\sum_{w=1}^W(x_{nchw}-\mu_{nc}(x))^2+\epsilon},
\end{align}
where $\epsilon$ is a small constent value to avoid zero division. IN has been used in style transfer and showed significant improvement~\cite{ulyanov2017improved}.

\noindent\textbf{Adative Instance Norm:}
AdaIN~\cite{huang2017arbitrary} extends the instance normalization into adaptive affine transformation in which the affine parameters are learned from the style image:
\begin{align} 
    \text{AdaIN}(x, y) = \sigma(y)\left(\frac{x-\mu(x)}{\sigma(x)}\right) + \mu(y),
\end{align}
where $y$ is a style image. AdaIN has been widely used in style transfer~\cite{liu2019few,lee2018diverse} due to its excellent performance.

\noindent\textbf{Additive Instance Norm:}
With the motivation that in text-to-image synthesis the noise and the sentence embedding should be decomposed, which is similar to the "content" and "style" decomposition in style transfer, it is intuitive to apply AdaIN in text-to-image synthesis. However, the condition and output in style transfer are in the same modality~(image), while in text-to-image synthesis, the condition and output are in different modalities~(text and image). So learning the parameters for AdaIN in text-to-image is more difficult because aligning the feature map between different modalities is more challenging. 

\begin{figure}[t]
    \centering
      \includegraphics[width=0.75\linewidth]{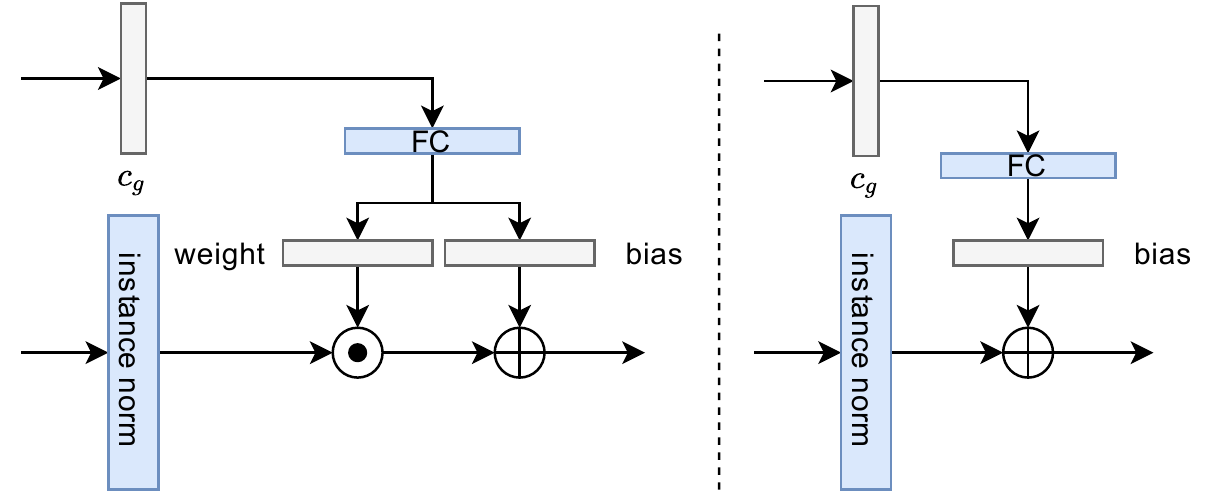}
  \caption{Left: adaptive instance norm~(AdaIN). Right: additive insrance norm~(AddIN).}
  \label{fig:adain_addin}
  \vspace{-2mm}
\end{figure}

In AdaIN, the weight and bias are two learnable parameters, however, if the cross-modality feature maps are not aligned, they will mislead the generation. When part of the feature map is mismatched, we can reduce the error in the bias by stacking multiple convolution blocks with the learnable normalization layer. But the error in the weight is more likely cumulative. So as shown in Fig.~\ref{fig:adain_addin}, we propose the \textbf{Add}itive \textbf{I}nstance \textbf{N}ormalization~(AddIN) which only modulates the feature with learnable bias to avoid the error accumulation:
\begin{align} 
    \text{AddIN}(x, c) = \frac{x-\mu(x)}{\sigma(x)} + T_c(c),
\end{align}
where $c$ is a condition feature, $T_c$ is a transformation to align the conditional feature map $c$ to $x$. 

\subsection{Factor Decomposed Base Generator}
We decompose the sentence embedding and the noise vector in the Factor Decomposed Base Generator~(FDBG).
Typically, given a noise vector $z$, which is sampled from $N(0,1)$, and a global condition vector $c_g$, we will concatenate them and synthesize the image from the concatenated feature. Here we propose to decompose these two factors. The global condition is used globally by applying the AddIN in the upsampling blocks. Applying AddIN in multiple upsampling blocks can also enhance the cross-modality semantic manipulation and reduce the adverse impact due to the cross-modality feature mismatching. 
Specifically, given the sampled noise vector $z$, global condition $c_g$:
\begin{align} 
    h_0^0 &= f_{re}(z),\\
    h_0^i & = U_i^{AddIN}(h_0^{i-1}, \bar{c_g}), \\
    h_0 &= h_0^4,\\
    \bar{c_g} &= T_{GC}(g_c),
\end{align}
where $U_i^{AddIN}$ is an upsampling block with AddIN, $f_{re}$ is a fully connected layer with reshape, $T_{GC}$ is a transformation block for the global condition to align the cross-modality features, which typically contains two fully connected layers, $i=1,2,3,4$ indicates that we use four upsampling blocks in the base generator.

\subsection{Attention Enhanced Next Generators}
The Attention Enhanced Next Generators~(AENG) is the same as the backbone~\cite{xu2018attngan} due to its excellent performance to refine the local texture by cross-modality attention. As illustrated in Fig.~\ref{fig:framework}, given the output of FDBG $h_0$, the local condition $c_l$, AENG refines the texture via the help of the  cross-modality attention module $F_i^{Attn}, i=1,2,\dots,m-1$, where $m$ is the stage number in the multi-stage generator. The synthesized images with increasing resolutions can be obtained:
\begin{align}
    \hat{x}_0 &= G_0(h_0), \\
    h_i &= F^i(h_{i-1}, F_i^{Attn}(h_{i-1}, c_l)),\\
    \hat{x}_i &= G_i(h_i), 
\end{align}
where $F^i$ is the feed-forward network for stage $i$ in the generator, which typically contains two residual blocks and an upsampling block, as illustrated in Fig.~\ref{fig:framework}. More details about the cross-modality attention can be found in~\cite{xu2018attngan}.

\subsection{Factor Decomposed Jointly Discriminator}
Typically, the conditional discriminator firstly extracts a 3-dimensional feature with a small spatial size from the input image via several downsampling blocks. Then concatenating this feature with the conditional feature to get the conditional prediction. Based on the aforementioned discussion about AddIN in Sec.~\ref{sec:addin}, we design a paired Factor Decomposed Jointly Discriminator~(FDJD) to modulate the image feature in the norm layer, as illustrated in Fig.~\ref{fig:framework} (b). The global condition $c_g$ is transformed by $T_{GC}$ before it is used as the bias to the AddIN in a convolutional block. Both the conditional and unconditional distributions are modeled in FDJD.

\section{Experimental Results}
We conduct extensive experiments to evaluate our proposed FDGAN. We firstly compare our proposed FDGAN with two recent state-of-the-arts~\cite{li2019controllable,zhu2019dm}, which have released source code, on the effectiveness and the efficiency. Then we analyze the different components in FDGAN to show that how they contribute to performance improvement. Finally, qualitative results are shown to verify the visual quality improvement.
\subsection{Datasets and Metrics}

\begin{table}
    \centering
    \caption{Training/testing set of CUB 200 and MS COCO datasets.}
    \label{tab:dataset_info}
    \begin{tabular}{l|c|c|c|c}
    \hline 
    \multicolumn{2}{c|}{Dataset} & Training & Testing & Total\\
    \hline
    \multirow{2}{*}{CUB 200} & class & 150 & 50 & 200 \\
     &image & 8855 & 2933 & 11788 \\
    \hline
    \multirow{2}{*}{MS COCO}& class & - & - & - \\
     &image & 82783 & 40470 & 123253  \\
    \hline
    \end{tabular}
    \vspace{-2mm}
\end{table}

\begin{table*}
    \centering
    \caption{Comparison with state-of-the-arts on CUB 200.}
    \label{tab:comparison_CUB}
    \begin{threeparttable}
    \begin{tabular}{ccccccc}
    \hline
     & \multicolumn{6}{c}{CUB 200}  \\
    \cline{2-7}
     Model&IS$\uparrow$&FID$\downarrow$&$R_1\uparrow$&Param&$T_{tr}$&$T_{te}$  \\
    \hline
    AttnGAN & 4.36 & 23.98 & 67.82 & 28.49 & 49.20 & 22.96 \\
    ControlGAN & 4.58 & - & 69.33 & 112.1 & 57.15 & 29.31 \\
    DM-GAN & 4.75 & 16.09 & 72.31 & 84.09 & 38.98 & 27.09 \\
    \hline
    
    FDGAN~(Ours)   & 4.44  &  {19.86} & {69.32} & {19.28} & 48.66 & 25.68 \\
    \hline
    \end{tabular}
    \begin{tablenotes}
        \item - $R_1$~(\%) denotes the Top-1 R precision. Param~(Mb) denotes the parameter size. $T_{tr}/T_{te}$ (ms) denotes training/testing time over the whole training/testing set averaged on a sample, respectively.
    \end{tablenotes}
    \end{threeparttable}
    \vspace{-2mm}
\end{table*}

\begin{table*}
    \centering
    \caption{Comparison with state-of-the-arts on CUB 200 and MS COCO.}
    \label{tab:comparison_coco}
    \begin{threeparttable}
    \begin{tabular}{ccccccc}
    \hline
     & \multicolumn{6}{c}{MS COCO}  \\
    \cline{2-7}
     Model&IS$\uparrow$&FID$\downarrow$&$R_1\uparrow$&Param&$T_{tr}$&$T_{te}$  \\
    \hline
    AttnGAN & 25.89 & 35.49 & 85.47 & 52.91 & 90.92 &  30.51 \\
    ControlGAN & 24.06 & - & 82.43 & 147.2  & 90.06 & 65.22 \\
    DM-GAN & 30.49 & 32.64 & 88.56 & 87.56 & 40.55 & 47.12 \\
    \hline
    
    FDGAN~(Ours)   & 4.44  &  {19.86} & {69.32} & {19.28} & 48.66 & 25.68 \\
    \hline
    \end{tabular}
    \begin{tablenotes}
        \item - $R_1$~(\%) denotes the Top-1 R precision. Param~(Mb) denotes the parameter size. $T_{tr}/T_{te}$ (ms) denotes training/testing time over the whole training/testing set averaged on a sample, respectively.
    \end{tablenotes}
    \end{threeparttable}
    \vspace{-2mm}
\end{table*}

\noindent\textbf{Dataset.} To evaluate the effectiveness of FDGAN, we use two widely used datasets: Caltech-UCSD Birds-200-2011~(CUB 200)~\cite{wah2011caltech} and Microsoft COCO~(MS COCO)~\cite{lin2014microsoft}, which are also been used in many previous works~\cite{xu2018attngan,yin2019semantics,tan2019semantics,zhu2019dm,li2019controllable}. CUB 200/MS COCO has 10/5 captions for each image, respectively.
More information about the two datasets is shown in Table~\ref{tab:dataset_info}. Training/testing split and preprocessing for the images are the same as that in~\cite{xu2018attngan,zhu2019dm}.

\noindent\textbf{Metrics.} Following the prior works on text-to-image synthesis~\cite{zhang2018photographic,tan2019semantics}, we use inception score~(IS)~\cite{salimans2016improved}, Fr\'echet inception distance~(FID)~\cite{heusel2017gans} and R precision~\cite{xu2018attngan} to evaluate the synthesis results. IS uses a pretrained Inception V3 model~\cite{szegedy2016rethinking} to measure both the objectiveness and the diversity of the synthesized images. A higher IS score means higher diversity of the synthesized images and the generated images are more likely to belong to one of the classes. FID measures the Fr\'echet distance between the real images and the generated images based on the Inception V3 features. A lower FID score means the generated image distribution is closer to the real image distribution. The R precision is obtained by retrieving the paired text given a generated image query. A higher R precision indicates better cross-modality semantic alignment between the text and the generated images.  Additionally, to compare the efficiency, we also give the parameter size~(the size of the saved model), and the training/testing computational complexity~(the averaged time for training/testing a sample).

\subsection{Implementation Details}
The only parameter which needs to be tuned is the size of the hidden layer in $T_{GC}$. Without the loss of generality, we set it as the mean of the input and output layer size. Other settings, including the preprocessing for the data, the optimizer, and the parameter averaging, follow the implementation in AttnGAN~\cite{xu2018attngan}.
We train our model on a V100 with 16G graphics memory. The training on CUB 200/MS COCO for 600/120 epochs costs about 3/10 GPU days, respectively.
The performance metrics, including IS, FID, and R precision, We follow the implementation of in DM-GAN\footnote{https://github.com/MinfengZhu/DM-GAN}~\cite{zhu2019dm}. The parameter size~(Mb) of the model is obtained by checking the size of the saved Pytorch\footnote{https://pytorch.org/} model.  The time complexity~(milliseconds) is calculated by training/testing on the whole training/testing set on a 2080TI GPU with 11G graphics memory and averaging the time on a single sample.
The code is avaliable in the supplementary materials.

\subsection{Comparison with the state-of-the-arts}
We compare our FDGAN with three related models. (1) AttnGAN is the baseline; (2) ControlGAN improved the AttnGAN via channel-wise attention and word-level discriminator; (3) DM-GAN added a dynamic memory module in every refined generator.  Both ControlGAN and DM-GAN improved the performance based on AttnGAN, but the much more parameters are used to achieve the improvement, and the training/testing computation complexity also increases considerably, as shown in Table~\ref{tab:comparison_CUB} and Table~\ref{tab:comparison_coco}.

Compared with AttnGAN, FDGAN can achieve better performance on both CUB 200 and MS COCO, on all the metrics. Meanwhile, we reduce the parameter size significantly. \textit{As far as we know, FDGAN is the first model that achieves better performance while fewer parameters are needed, compared with AttnGAN}. 

Compared with ControlGAN, FDGAN is competitive on CUB 200 and better on MS COCO, as shown in Table~\ref{tab:comparison_CUB} and Table~\ref{tab:comparison_coco}. Moreover, the parameters used in our FDGAN are much fewer than ControlGAN, and FDGAN is much more efficient than ControlGAN, too.
Compared with DM-GAN, FDGAN is not as excellent as DM-GAN on either CUB 200 or MS COCO, except that FDGAN is more efficient. However, the dynamic memory module used in DM-GAN is applied in the refined generators, which is compatible with our FDGAN. We can boost our model further by adding the dynamic memory module, but it is beyond the scope of this paper.

\begin{table}
    \centering
    \caption{Component Analysis for FDGAN on CUB 200.}
    \label{tab:component_analysis}
    \begin{threeparttable}
    \begin{tabular}{cccc}
    \hline
     Model&IS$\uparrow$&FID$\downarrow$&$R_1\uparrow$ \\
    \hline
    Baseline~(AttnGAN) & 4.36$\pm$0.03 & 23.98 & 67.82\\
    \hline
    Baseline + AdaIN & 4.40$\pm$0.05 & 20.66 & 65.53   \\
    Baseline + FDBG  & 4.41$\pm$0.07 & 20.51 & 68.80  \\
    Baseline + FDJD  & \textbf{4.47$\pm$0.07} & 23.71 & 65.72 \\
    FDGAN~(Ours)   & 4.44$\pm$0.06  &  \textbf{19.86} & \textbf{69.32}     \\
    \hline
    \end{tabular}
\end{threeparttable}
\vspace{-2mm}
\end{table}

\begin{figure*}[t]
    \centering
      \includegraphics[width=0.99\linewidth]{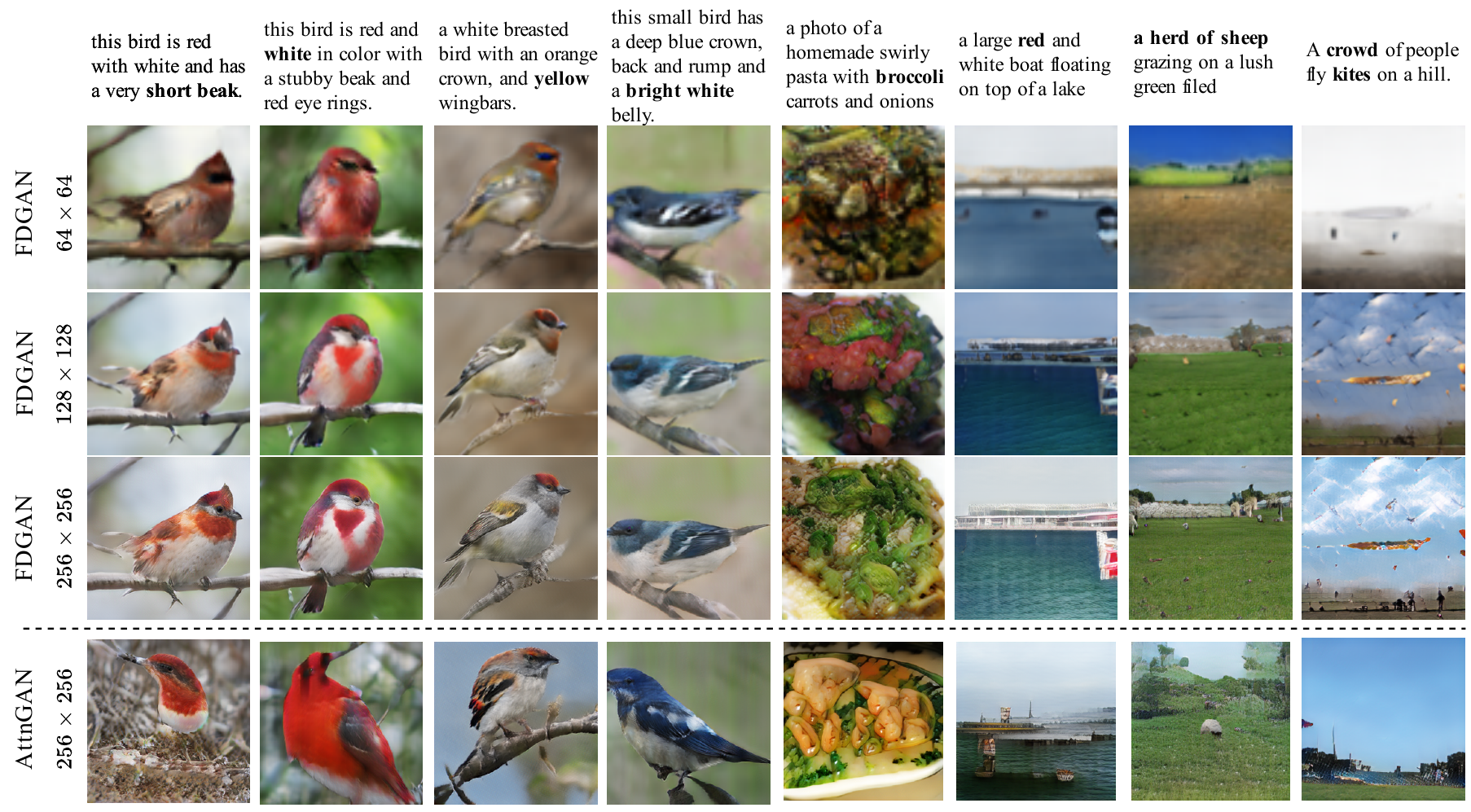}
  \caption{Qualitative results on CUB 200~(left 4 columns) and MS COCO~(right 4 columns) for our FDGAN~(ours, top 3 rows) and AttnGAN~(baseline, bottom 1 row). The words that our FDGAN captures while AttnGAN ignores are highlighted with boldface. (Best viewed in color. Zoom in to see the details.)}
  \label{fig:qualitative_result}
   \vspace{-2mm}
\end{figure*}
\begin{figure}[t]
    \centering
      \includegraphics[width=0.99\linewidth]{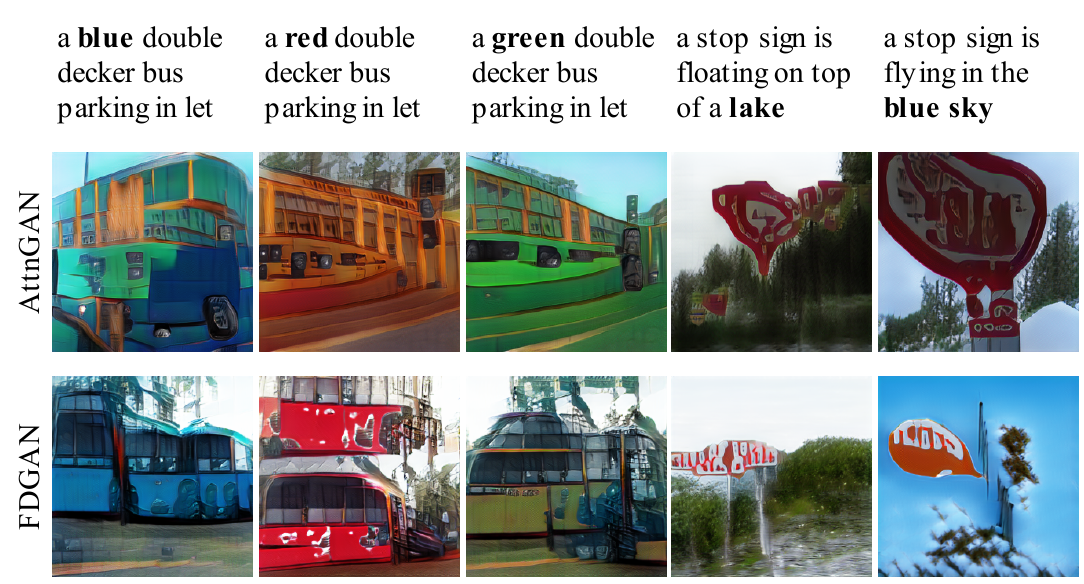}
  \caption{Example results of our FDGAN~(top) and AttnGAN~(bottom) on MS COCO when changing some words to show the generalization of our FDGAN. 
  We can see that our FDGAN can capture the semantic in the text better than AttnGAN when only changing one word in the text, indicating that FDGAN has better generalization.
   }
  \label{fig:fine_grained}
  \vspace{-2mm}
\end{figure}

\subsection{Component Analysis}
In this section, we want to answer the following two questions via component analysis: (1) how the Factor Decomposed Base Generator~(FDBG) and the Factor Decomposed Jointly Discriminator~(FDJD) contribute to FDGAN? (2) does additive instance norm~(AddIN) perform better than adaptive instance norm~(AdaIN)?

To answer the above two question, we train several models with the following 4 different settings: Baseline + AdaIN indicates that we replace the AddIN layer with the AdaIN layer for FDBG; Baseline + FDBG indicates that we add the FDBG module on the baseline; Baseline + FDJD indicates that we only use the FDJD on the baseline, the FDBG is not used; FDGAN indicates that our full model with both FDBG and FDJD.

As illustrated in Table~\ref{tab:component_analysis}, compared with the baseline~(1st row), Baseline + FDBG~(3rd row) achieve better IS, FID, and R precision, implying that FDBG can bring stable performance improvement, demonstrating that factor decomposition can help us get a better generative model for text-to-image synthesis. Compared with baseline, Baseline + FDJD~(4th row) only achieve better IS, while the FID and R precision is slightly dropped, while using FDBG and FDJD in pairs in FDGAN~(5th row) can boost the performance further, implying that \textit{using FDBG and FDJD in pairs can bring considerable, stable performance improvement}. These results demonstrate the effectiveness of our motivation: \textit{factor decomposition in text-to-image synthesis can lead performance improvement}.

Compared with the baseline~(1st row), Baseline + AdaIN~(2nd row) can achieve better IS and FID, while the R precision drops slightly~(from 67.82\% to 65.53\%), indicating that AdaIN may affect the cross-modality semantic alignment and leads to a drop of R precision. By comparison, Baseline + FDBG~(3rd row), which uses AddIN, can achieve an R precision improvement~(from 67.82\% to 68.80\%) with comparable IS and FID with that of Baseline + AdaIN~(2nd row). This verifies the hypothesis that \textit{the weight in AdaIN may affect the cross-modality semantic alignment and AddIN can learn to align the text-image semantic alignment better}.
\subsection{Qualitative Results}
In addition to the quantitative comparison with the baseline, we also show the qualitative comparison results to show that our FDGAN can better capture the semantic in the text due to the factor decomposition. As shown in Fig.~\ref{fig:qualitative_result}, on both CUB 200 and MS COCO, our model can (1) increasingly refine the generated image to align the semantic gradually; (2) capture the semantic better than the baseline~(AttnGAN)~(\textit{e.g.} the words about color in CUB 200 and the objects in MS COCO, as highlighted in Fig.~\ref{fig:qualitative_result}). Additionally, we test our FDGAN on some new sentence to show that FDGAN is with better generalization for the unseen text, as illustrated in Fig.~\ref{fig:fine_grained}.

\section{Conclusion}
In this paper, we attempted to decompose the factors~(sentence embedding and noise) for text-to-image synthesis and proposed Factor Decomposed Generative Adversarial Networks~(FDGAN). We also design an additive instance norm~(AddIN) layer to apply the sentence embedding into the generator and discriminator. The experimental results show that (1) by decomposing the noise and the global condition, the latent factors in the generated image can be disentangled; 
(2) our proposed FDGAN can help capture the semantic and achieve better performance with fewer parameters; 
(3) our proposed AddIN can help synthesize images with better semantic consistency.


\bibliography{FDGAN}

\begin{thebibliography}{10}
\expandafter\ifx\csname url\endcsname\relax
  \def\url#1{\texttt{#1}}\fi
\expandafter\ifx\csname urlprefix\endcsname\relax\def\urlprefix{URL }\fi
\expandafter\ifx\csname href\endcsname\relax
  \def\href#1#2{#2} \def\path#1{#1}\fi

\bibitem{reed2016generative}
S.~Reed, Z.~Akata, X.~Yan, L.~Logeswaran, B.~Schiele, H.~Lee, Generative
  adversarial text to image synthesis, in: Proceedings of ICML, 2016, pp.
  1060--1069.

\bibitem{goodfellow2014generative}
I.~Goodfellow, J.~Pouget-Abadie, M.~Mirza, B.~Xu, D.~Warde-Farley, S.~Ozair,
  A.~Courville, Y.~Bengio, Generative adversarial nets, Proceedings of NeurIPS
  27 (2014) 2672--2680.

\bibitem{mirza2014conditional}
M.~Mirza, S.~Osindero, Conditional generative adversarial nets, arXiv preprint
  arXiv:1411.1784 (2014).

\bibitem{zhang2017stackgan}
H.~Zhang, T.~Xu, H.~Li, S.~Zhang, X.~Wang, X.~Huang, D.~N. Metaxas, Stackgan:
  Text to photo-realistic image synthesis with stacked generative adversarial
  networks, in: Proceedings of ICCV, 2017, pp. 5907--5915.

\bibitem{zhang2018stackgan++}
H.~Zhang, T.~Xu, H.~Li, S.~Zhang, X.~Wang, X.~Huang, D.~N. Metaxas, Stackgan++:
  Realistic image synthesis with stacked generative adversarial networks, IEEE
  TPAMI 41~(8) (2018) 1947--1962.

\bibitem{xu2018attngan}
T.~Xu, P.~Zhang, Q.~Huang, H.~Zhang, Z.~Gan, X.~Huang, X.~He, Attngan:
  Fine-grained text to image generation with attentional generative adversarial
  networks, in: Proceedings of CVPR, 2018, pp. 1316--1324.

\bibitem{zhu2019dm}
M.~Zhu, P.~Pan, W.~Chen, Y.~Yang, Dm-gan: Dynamic memory generative adversarial
  networks for text-to-image synthesis, in: Proceedings of CVPR, 2019, pp.
  5802--5810.

\bibitem{tulyakov2018mocogan}
S.~Tulyakov, M.-Y. Liu, X.~Yang, J.~Kautz, Mocogan: Decomposing motion and
  content for video generation, in: Proceedings of CVPR, 2018, pp. 1526--1535.

\bibitem{huang2017arbitrary}
X.~Huang, S.~Belongie, Arbitrary style transfer in real-time with adaptive
  instance normalization, in: Proceedings of ICCV, 2017, pp. 1501--1510.

\bibitem{liu2019few}
M.-Y. Liu, X.~Huang, A.~Mallya, T.~Karras, T.~Aila, J.~Lehtinen, J.~Kautz,
  Few-shot unsupervised image-to-image translation, in: Proceedings of CVPR,
  2019, pp. 10551--10560.

\bibitem{lee2018diverse}
H.-Y. Lee, H.-Y. Tseng, J.-B. Huang, M.~Singh, M.-H. Yang, Diverse
  image-to-image translation via disentangled representations, in: Proceedings
  of ECCV, 2018, pp. 35--51.

\bibitem{qiao2019mirrorgan}
T.~Qiao, J.~Zhang, D.~Xu, D.~Tao, Mirrorgan: Learning text-to-image generation
  by redescription, in: Proceedings of CVPR, 2019, pp. 1505--1514.

\bibitem{tan2019semantics}
H.~Tan, X.~Liu, X.~Li, Y.~Zhang, B.~Yin, Semantics-enhanced adversarial nets
  for text-to-image synthesis, in: Proceedings of ICCV, 2019, pp. 10501--10510.

\bibitem{yin2019semantics}
G.~Yin, B.~Liu, L.~Sheng, N.~Yu, X.~Wang, J.~Shao, Semantics disentangling for
  text-to-image generation, in: Proceedings of CVPR, 2019, pp. 2327--2336.

\bibitem{cheng2020rifegan}
J.~Cheng, F.~Wu, Y.~Tian, L.~Wang, D.~Tao, Rifegan: Rich feature generation for
  text-to-image synthesis from prior knowledge, in: Proceedings of CVPR, 2020,
  pp. 10911--10920.

\bibitem{dumoulin2016learned}
V.~Dumoulin, J.~Shlens, M.~Kudlur, A learned representation for artistic style,
  Proceedings of ICLR (2017) 1--7.

\bibitem{ulyanov2017improved}
D.~Ulyanov, A.~Vedaldi, V.~Lempitsky, Improved texture networks: Maximizing
  quality and diversity in feed-forward stylization and texture synthesis, in:
  Proceedings of CVPR, 2017, pp. 6924--6932.

\bibitem{li2019controllable}
B.~Li, X.~Qi, T.~Lukasiewicz, P.~Torr, Controllable text-to-image generation,
  in: Proceedings of of NeurIPS, 2019, pp. 2065--2075.

\bibitem{wah2011caltech}
C.~Wah, S.~Branson, P.~Welinder, P.~Perona, S.~Belongie, The caltech-ucsd
  birds-200-2011 dataset (2011).

\bibitem{lin2014microsoft}
T.-Y. Lin, M.~Maire, S.~Belongie, J.~Hays, P.~Perona, D.~Ramanan,
  P.~Doll{\'a}r, C.~L. Zitnick, Microsoft coco: Common objects in context, in:
  Proceedings of ECCV, Springer, 2014, pp. 740--755.

\bibitem{zhang2018photographic}
Z.~Zhang, Y.~Xie, L.~Yang, Photographic text-to-image synthesis with a
  hierarchically-nested adversarial network, in: Proceedings of CVPR, 2018, pp.
  6199--6208.

\bibitem{salimans2016improved}
T.~Salimans, I.~Goodfellow, W.~Zaremba, V.~Cheung, A.~Radford, X.~Chen,
  Improved techniques for training gans, in: Proceedings of NeurIPS, 2016, pp.
  2234--2242.

\bibitem{heusel2017gans}
M.~Heusel, H.~Ramsauer, T.~Unterthiner, B.~Nessler, S.~Hochreiter, Gans trained
  by a two time-scale update rule converge to a local nash equilibrium, in:
  Proceedings of NeurIPS, 2017, pp. 6626--6637.

\bibitem{szegedy2016rethinking}
C.~Szegedy, V.~Vanhoucke, S.~Ioffe, J.~Shlens, Z.~Wojna, Rethinking the
  inception architecture for computer vision, in: Proceedings of CVPR, 2016,
  pp. 2818--2826.

\end{thebibliography}

\end{document}